\def\papertitle{Arbitrary Polygon Oscillator: Generalizing Polygonal Synthesis to Arbitrary Shapes, Morphing, and Three-Dimensional Polyhedra}
\def\paperauthorA{Antonio Argentieri}
\def\paperauthorB{Francesco Scagliola}
\documentclass[twoside,a4paper]{article}
\usepackage{etoolbox}
\usepackage[print]{dafx26v3}

\usepackage{amsmath,amssymb,amsfonts,amsthm}
\usepackage{siunitx}
\usepackage{euscript}
\usepackage[T1]{fontenc}
\usepackage[utf8]{inputenc}
\usepackage{ifpdf}
\usepackage[english]{babel}
\usepackage{caption}
\usepackage{subfig}
\usepackage{color}
\usepackage{booktabs}
\usepackage{enumitem}
\usepackage{xcolor}
\usepackage[stretch=0,shrink=20]{microtype}

\input glyphtounicode
\ninept

\newcounter{numauth}
\newcounter{listcnt}
\newcommand\authcnt[1]{\ifdefined#1 \stepcounter{numauth} \fi}
\newcommand\addauth[1]{
\ifdefined#1
\stepcounter{listcnt}
\ifnum \value{listcnt}<\value{numauth}
\appto\authorslist{, #1}
\else
\appto\authorslist{~and~#1}
\fi
\fi}
\authcnt{\paperauthorB}
\authcnt{\paperauthorC}
\authcnt{\paperauthorD}
\authcnt{\paperauthorE}
\authcnt{\paperauthorF}
\authcnt{\paperauthorG}
\authcnt{\paperauthorH}
\authcnt{\paperauthorI}
\authcnt{\paperauthorJ}
\def\authorslist{\paperauthorA}
\addauth{\paperauthorB}
\addauth{\paperauthorC}
\addauth{\paperauthorD}
\addauth{\paperauthorE}
\addauth{\paperauthorF}
\addauth{\paperauthorG}
\addauth{\paperauthorH}
\addauth{\paperauthorI}
\addauth{\paperauthorJ}

\usepackage{times}

\newif\ifpdf
\ifx\pdfoutput\relax
\else
   \ifcase\pdfoutput
      \pdffalse
   \else
      \pdftrue
   \fi
\fi

\ifpdf
  \usepackage[pdftex,
    pdftitle={\papertitle},
    pdfauthor={Antonio Argentieri and Francesco Scagliola},
    pdfsubject={Proceedings of the 29th International Conference on Digital Audio Effects (DAFx26)},
    colorlinks=false,
    bookmarksnumbered,
    pdfstartview=XYZ
  ]{hyperref}
  \usepackage[pdftex]{graphicx}
\else
  \usepackage[dvips]{epsfig,graphicx}
  \usepackage[dvips,
    pdftitle={\papertitle},
    pdfauthor={Antonio Argentieri and Francesco Scagliola},
    pdfsubject={Proceedings of the 29th International Conference on Digital Audio Effects (DAFx26)},
    colorlinks=false,
    bookmarksnumbered,
    pdfstartview=XYZ
  ]{hyperref}
\fi
\usepackage[hypcap=true]{caption}
\title{\papertitle}

\affiliation
{\paperauthorA \ and \paperauthorB}
{Conservatorio Niccol\`o Piccinni di Bari\\
Bari, Italy\\
{\tt \href{mailto:antonioargentieri76@gmail.com}{antonioargentieri76@gmail.com} , \href{mailto:francesco.scagliola@gmail.com}{francesco.scagliola@gmail.com}}
}

\begin{document}
\ifpdf
  \DeclareGraphicsExtensions{.png,.jpg,.pdf}
\else
  \DeclareGraphicsExtensions{.eps}
\fi

\maketitle

\begin{abstract}
Polygonal synthesis generates audio by traversing the perimeter of a polygon
with a phasor~\cite{Hohnerlein:2016}; prior work uses a constant angular velocity,
whereas the proposed system adopts constant arc-length (perimeter) velocity.
Existing formulations operate on regular, parametrically defined polygons,
producing smooth timbral transitions within a single family of shapes.
 This paper generalizes polygonal synthesis around a unified
arc-length engine: vertex data of any origin feed the same DSP pipeline.
First, we adapt the oscillator to accept arbitrary vertex configurations
from an external buffer, opening the possibility for a broad class of closed polygons~--- regular, irregular,
or star-shaped~--- to function as a waveform generator. Second, a hybrid interpolation algorithm enables smooth morphing
between polygons with unequal vertex counts, passing through intermediate
shapes that have no parametric description. Third, we extend the paradigm to
three dimensions: a convex polyhedron rotated about three axes is sliced by a
fixed horizontal plane, and the resulting cross-section yields a continuously
variable polygon controlled by the solid's orientation.
The system runs in RNBO (Cycling~'74) with a geometry caching strategy that
avoids per-sample recomputation. Antialiasing combines a four-point polyBLAMP
correction derived from runtime Bézier tangents with adaptive oversampling,
adapting the correction geometrically to general vertex configurations without
per-shape analytical derivation.
\end{abstract}

\section{Introduction}
\label{sec:intro}

The idea that a polygon traversed by a phasor generates a musically useful
waveform was introduced by Chapman~\cite{Chapman:2014}, who established the
formal relationship between regular polygon geometry and harmonic content,
and independently formalized by Hohnerlein et al.~\cite{Hohnerlein:2016},
who defined the synthesis as a phasor sampling a variable polygon in polar
space at constant angular velocity.
Their contributions define the field now commonly referred to as polygonal
synthesis. Chapman's work explored regular and star polygons described by
discrete Schl\"{a}fli symbols $\{n/q\}$; Hohnerlein et al.'s key innovation
was a \emph{continuous} polygon order parameter~$n$, enabling smooth
timbral transitions sweeping through triangle, square, pentagon and beyond.

The present paper takes a different approach: rather than parameterizing the
shape mathematically, we define it by its vertices. A vertex buffer provides
complete freedom of form~--- any closed polygon, regular or irregular, convex
or concave, can function as a waveform generator. The interpolation between
shapes is then not a matter of changing a parameter but of navigating between
two drawn configurations, passing through intermediate forms that have no
parametric description. This opens a different region of the synthesis space,
one that includes shapes unreachable by any continuous order parameter.

\subsection{Contributions}
\label{ssec:contributions}

\begin{itemize}
  \item \textbf{Arbitrary polygon oscillator}: closed polygons of general shape 
    specified by an external vertex buffer, traversed via arc-length parameterization
    for constant perimeter velocity.
  \item \textbf{Hybrid shape interpolation}: smooth morphing between polygons
    with different vertex counts using angular correspondence for convex
    shapes and perimeter-based midpoint expansion for concave shapes, with
    vertex sharpness preserved throughout transitions.
   \item \textbf{Runtime poly\-BLAMP antialiasing}:  polyBLAMP correction, computed from geometric Bézier tangents at each vertex using four points, generalizes the closed-form derivative expressions of~\cite{parker2017} to arbitrary vertex configurations without per-shape analytical derivation.
   \item \textbf{3D extension via planar cross-section}: convex polyhedra
    inscribed in a sphere are traversed by a cutting plane whose orientation
    can be rotated freely in three axes; the resulting planar polygon provides
    a continuously variable waveform controlled by spatial orientation.
\end{itemize}

\section{Background}
\label{sec:background}

\subsection{Core Mechanism}
\label{ssec:bg_mechanism}
Polygonal synthesis produces audio by traversing the perimeter of a closed
polygon with a phasor and reading the resulting $x$ and $y$ coordinates as a
two-dimensional output. The traversal speed sets the fundamental
frequency~$f_0$; the geometry sets the waveform and hence the spectrum.
A circle at constant angular velocity gives a pure sine on each channel;
a triangle, square, or star introduces corners~--- abrupt changes in
traversal direction~--- that generate harmonics. Two aspects of the geometry
play distinct roles: the polygon's rotational symmetry fixes \emph{which}
harmonics are present, while the number and sharpness of its corners set
\emph{how} energy is distributed among them~--- more sides (for a regular
polygon, a higher symmetry order) approach a circle and attenuate the high
harmonics, fewer or sharper corners enrich the spectrum. This separation is
made precise in §3.5. The output is inherently two-dimensional~--- the $x(\phi)$
and $y(\phi)$ channels carry distinct waveforms whose phase relationship follows
the polygon's asymmetry~--- making the oscillator a native stereo source.

\begin{figure}[ht]
  \centering
  \includegraphics[width=1 \columnwidth]{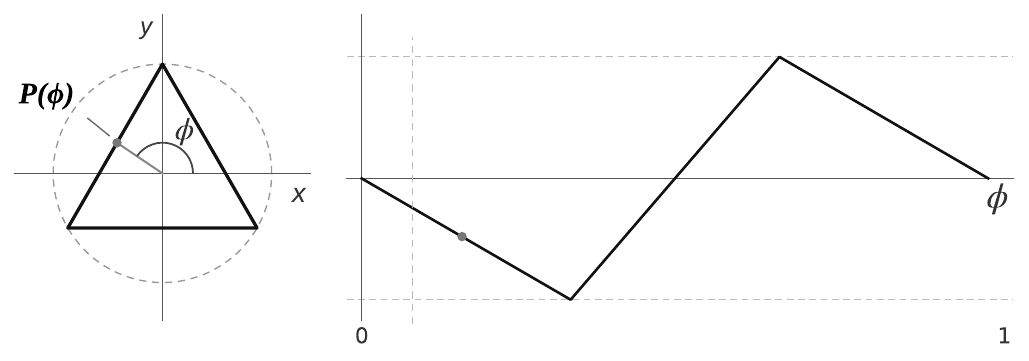}
  \caption{Polygonal synthesis illustrated for an equilateral triangle
    ($N=3$, $f_0=\SI{110}{\hertz}$). \emph{Left:} the phasor traverses
    the perimeter at constant velocity; the traversal point
    $\mathbf{P}(\phi)$ is the polygon coordinate at phase~$\phi$.
    \emph{Right:} the resulting output waveform (one coordinate of
    $\mathbf{P}(\phi)$) over one complete traversal, the phase~$\phi$
    running from $0$ to $1$.}
  \label{fig:bg_primer}
\end{figure}

\subsection{Relation to Other Synthesis Techniques}
\label{ssec:bg_related_synth}
The proposed system plays an arbitrary polygon and morphs it continuously
 into any other over a chosen duration.
As Hohnerlein et al.~\cite{Hohnerlein:2016} note, polygonal synthesis produces timbral
behaviour most closely analogous to digital waveshaping~\cite{LeBrun:1979}: both pass a periodic input
through a transfer function that shapes the spectrum by nonlinear distortion.
The paradigms differ structurally: waveshaping maps an amplitude signal $x = \cos\theta$
through a fixed scalar function $f(x)$, while polygonal synthesis maps a phase index
$\phi \in [0,1)$ through a geometric transfer function
$g(f,p_1,\dots,p_n)$ defined by vertex coordinate, yielding a native
two-dimensional output $(x(\phi),\, y(\phi))$.

 Polygonal synthesis also differs from wave terrain
synthesis~\cite{Mitsuhashi:1982}, which shares the 2-D orbit $(x(t),y(t))$ but
uses it to address a scalar function $F(x,y)$ for a single output; here the
orbit coordinates \emph{are} the output, no function is applied to the traversal
result. Similarly, this technique is unrelated to wavetable synthe-
sis, in which a phasor reads a fixed, time-invariant waveform from
a stored buffer: here the waveform is not stored but geometrically
reconstructed from vertex data, and it varies continuously during
morphing.

\subsection{The Angular-Velocity Baseline and Its Evolution}
\label{ssec:bg_sota}

Chapman~\cite{Chapman:2014} introduced $n$-gon waves as waveforms derived from
the geometry of regular polygons in the time domain, establishing the formal
relationship between polygon order and harmonic content, and showing that
sawtooth, triangle, and square waves are special cases of a broader family.
Star polygons described by Schl\"{a}fli symbols $\{n/q\}$ were also explored.

Hohnerlein, Rest, and Smith~\cite{Hohnerlein:2016} formalized polygonal
synthesis as a phasor rotating in the complex plane at constant angular
velocity~$\dot{\theta} = 2\pi f_0$, with the polygon's radial distance
function $r(\theta)$ determining the output. The key innovation was a
continuously variable polygon order~$n$, enabling smooth timbral transitions
between integer polygon orders. For a regular $N$-gon this approach produces a
sparse spectrum with energy only at harmonics $(kN \pm 1)f_0$ ($N$-fold symmetry; for a regular
N-gon the symmetry order coincides with the vertex count, a coincidence that does not hold for general shapes — see §3.5).
The harmonic amplitudes are determined by the time the phasor spends on each
portion of the perimeter, which under angular-velocity traversal is
proportional to the subtended angle rather than the edge length.
This work formed the basis for the polygogo Eurorack module~\cite{Polygogo:2019},
a commercial hardware implementation that popularized the technique.

Peschke and Berndt~\cite{Peschke:2017} proposed Cyclone, a geometric oscillator
that derives waveforms from cyclic Bézier paths traversed at constant arc-length
velocity, and outlined a polygon-based successor (Zykloid) that would support
freely drawn shapes and morphing. No formal publication or implementation of
Zykloid has subsequently appeared.

%

\section{Arbitrary Polygon Oscillator}
\label{sec:oscillator}

\subsection{Vertex Definition}
\label{ssec:vertexdef}

The oscillator accepts $N$ vertices $\{\mathbf{V}_i = (x_i, y_i)\}_{i=0}^{N-1}$ from
an external buffer. Scalars are italic ($x_i$, $\ell_i$); points and vectors are bold
($\mathbf{V}_i$, $\mathbf{P}$). All subsequent processing is performed relative to the centroid
$C = \bigl(\tfrac{1}{N}\sum_{i=0}^{N-1}x_i,\;\tfrac{1}{N}\sum_{i=0}^{N-1}y_i\bigr)$.
The signal processing pipeline proceeds in three stages: geometric
transformations are applied to the vertex coordinates
(Section~\ref{ssec:transforms}), the modified polygon is traversed via
arc-length parameterization (Section~\ref{ssec:traversal}), and the
traversal point $\mathbf{P}(\phi) = (x(\phi), y(\phi))$ is delivered as a
two-dimensional coordinate signal available for further processing.


\subsection{Geometric Transformations}
\label{ssec:transforms}

The following transformations are offered as real-time sound design controls;
they operate on vertex coordinates relative to the centroid before traversal,
and apply uniformly to all polygon types.

\textbf{Rotation.} The standard  rotation matrix $R(\theta)$
is applied uniformly to all vertices. Unlike a
traversal phase offset (inaudible when static), it mixes the two
channels, reshaping each waveform and the stereo image; swept, it
animates the stereo field.

\textbf{Squeeze.} Scales $x$ by $(1{+}s/2)$ and $y$ by $(1{-}s/2)$,
breaking symmetry; $s\in[-1,1]$ is a non-degenerate $\pm50\%$ window
(an axis collapses only at $|s|{=}2$), not area-preserving
($\det=1-s^2/4$).
\subsubsection{Edge Curvature}
\label{ssec:curvature}

Edges can be curved by displacing a quadratic B\'{e}zier control point from
the edge midpoint along the unit vector perpendicular to the edge:
\begin{equation}
  \mathbf{P}_{c,i} = \mathbf{m}_i + \hat{\mathbf{n}}_i \cdot \kappa \cdot \frac{\|\mathbf{V}_{i+1} - \mathbf{V}_i\|}{2}
  \label{eq:curvature}
\end{equation}
where $\mathbf{m}_i = (\mathbf{V}_i + \mathbf{V}_{i+1})/2$ is the midpoint of edge $i$,
$\hat{\mathbf{n}}_i$ is the unit vector obtained by rotating the edge
direction $90°$ counter-clockwise:
\begin{equation*}
  \hat{\mathbf{n}}_i = \frac{1}{\|\mathbf{V}_{i+1}-\mathbf{V}_i\|}\bigl(-(y_{i+1}-y_i),\;x_{i+1}-x_i\bigr),
\end{equation*}
$\kappa$ is the edge curvature parameter. Negative $\kappa$ bows
edges outward (rounder shape, smoother waveform); positive $\kappa$ bows them
inward (concave sides, sharper waveform).
At fixed vertex count, $\kappa$ keeps the harmonic positions and only redistributes their magnitudes~--- concave curvature brightening the tone, convex curvature rounding it toward a sine (companion page\footnote{\url{https://www.antonioargentieri.com/polygon_demo/}}).

\subsection{Arc-Length Traversal}
\label{ssec:traversal}

 With arbitrary vertex configurations, edges vary in length. Allocating equal
phase to each edge would cause the phasor to traverse short edges slowly and
long edges quickly, introducing a periodic pitch fluctuation audible as
timbral instability on polygons with unequal sides. Arc-length
parameterization eliminates this artefact by maintaining constant perimeter
velocity, so that the phasor spends time on each edge in proportion to its
length.

 The total perimeter $L = \sum_{i=0}^{N-1} \ell_i$ is computed once per
geometry update, and the phase $\phi \in [0,1)$ is mapped to arc-length
distance $s = \phi \cdot L$. Given $s$, the containing edge is found by
accumulating lengths, and the local parameter $u \in [0,1]$ is
computed as:
\begin{equation}
  u = \frac{s - \sum_{j=0}^{i-1}\ell_j}{\ell_i}
  \label{eq:tlocal}
\end{equation}
The traversal point is then obtained by evaluating the quadratic
B\'{e}zier curve along the edge:
\begin{equation}
  \mathbf{P}(u) = (1-u)^2\,\mathbf{V}_i + 2(1-u)\,u\,\mathbf{P}_{c,i}
                + u^2\,\mathbf{V}_{i+1}
  \label{eq:bezier_traversal}
\end{equation}
 where $\mathbf{P}_{c,i}$ is the Bézier control point of edge $i$
(Section~\ref{ssec:curvature}). For $\kappa=0$,
$\mathbf{P}_{c,i} = (\mathbf{V}_i + \mathbf{V}_{i+1})/2$
and~\eqref{eq:bezier_traversal} reduces to linear interpolation.
The $x$ and $y$ components of $\mathbf{P}(u)$ are delivered directly as the
two native audio output channels; constant-velocity traversal ensures that
the oscillation frequency is determined solely by the phasor rate,
independently of the polygon's shape or edge distribution.

 Since the exact arc length of a quadratic Bézier curve requires an elliptic
integral with no closed-form solution at runtime, each $\ell_i$ is estimated
via Gravesen's convex combination of the chord length and the
control-polygon length~\cite{gravesen1997}:
\begin{equation}
  \ell_i = \frac{2\,\|\mathbf{V}_{i+1}-\mathbf{V}_i\|
              + \|\mathbf{P}_{c,i}-\mathbf{V}_i\|
              + \|\mathbf{V}_{i+1}-\mathbf{P}_{c,i}\|}{3}
  \label{eq:edgelen}
\end{equation}
 This approximation achieves better than $0.1\%$ relative error for the
curvature range used here~\cite{gravesen1997}, reduces to the Euclidean
chord when $\kappa=0$, and is consistent with the chord directions used
in the polyBLAMP \cite{parker2017} correction (Section~\ref{sec:aa}). Edge geometry is cached
and recomputed only when the geometry changes, avoiding per-sample
recalculation.

\subsection{Comparison with Angular-Velocity Traversal}
\label{ssec:traversal_compare}

 As established in Section~\ref{ssec:bg_sota}, arc-length and angular-velocity
traversal both confine spectral energy to harmonics $(kN \pm 1)f_0$ for regular
$N$-gons, but produce different harmonic amplitudes because the time the
phasor spends on each portion of the perimeter differs between the two
methods. Two independent mechanisms allow the arc-length output to
approximate the timbral character of the angular-velocity baseline.

The first is edge curvature (Section~\ref{ssec:curvature}): the
angular-velocity phasor implicitly traces curved paths between vertices
because it spends more time on distant edges, producing rounded waveform
transitions. With zero curvature, arc-length traversal instead produces 
piecewise-linear transitions.
 The second is harmonic balance. Weighting the output channels by
$\|\mathbf{P}(\phi)\| / r_\text{max}$, where
$r_\text{max} = \max_{\phi} \|\mathbf{P}(\phi)\|$ is the maximum distance from
the centroid to any point on the perimeter (not merely to a vertex, 
since Bézier edge curvature can place the farthest point along
the arc rather than at a corner), encodes the polygon's radial profile along the arc-length trajectory rather than its cartesian
projection. As $\|\mathbf{P}(\phi)\|$ shares the fundamental period of
$(x(\phi),y(\phi))$, this weighting does not introduce new spectral components:
 it redistributes energy across existing harmonics, shifting the timbre 
toward a radial reading of the polygon~--- precisely what angular-velocity traversal produces.

\begin{figure}[ht]
  \centering
  \includegraphics[width=1\columnwidth]{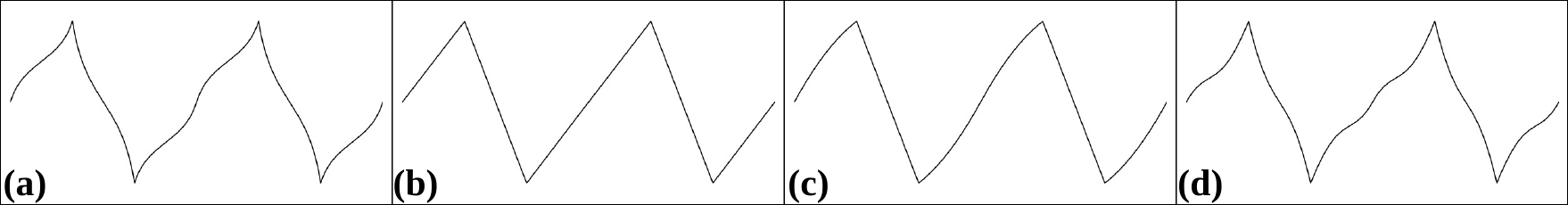}\\[3pt]
  \includegraphics[width=1.\columnwidth]{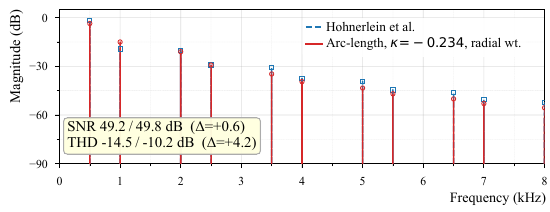}
  \caption{Triangle, $N=3$. \textbf{(top)}~Waveforms
    ($f_0=\SI{110}{\hertz}$, $f_s=\SI{44.1}{\kilo\hertz}$):
    (a)~angular-velocity~\cite{Hohnerlein:2016}, (b)~arc-length, no correction,
    (c)~$\kappa=-0.234$, (d)~$\kappa=-0.234$ with radial weighting.
    \textbf{(bottom)}~Spectra of (a) (blue dashed) vs (d) (red) at
    $f_0=\SI{500}{\hertz}$, $f_s=\SI{48}{\kilo\hertz}$; SNR/THD deltas inset.}
  \label{fig:traversal_comparison}
\end{figure}

Figure~\ref{fig:traversal_comparison} shows the progressive visual approximation toward 
the angular-velocity output for $N=3$: the uncorrected arc-length output (b) shows piecewise-linear transitions and a
different harmonic balance relative to the Hohnerlein reference (a); edge curvature
$\kappa=-0.234$ (c), empirically chosen to minimise the visual discrepancy, rounds the waveform 
transitions; adding radial weighting (d) further aligns the harmonic balance, producing
a waveform visually similar to the angular-velocity output.

\subsection{Spectral Characterization by Symmetry}
\label{ssec:symmetry}

The harmonic content of the output is determined not by the vertex count~$N$
but by the \emph{rotational symmetry order}~$M$: the largest integer $M\ge1$
such that a rotation of $2\pi/M$ maps the polygon onto itself.
For a regular $N$-gon, $M=N$; for a five-pointed star drawn as a concave
10-vertex polygon, $M=5$ regardless of vertex count; for a polygon with no
rotational symmetry, $M=1$.

A polygon with $M$-fold symmetry has $M$ identical sectors, each spanning
$1/M$ of the total perimeter and rotated by $2\pi/M$ relative to the
previous one; under arc-length traversal, advancing the phase by $1/M$
therefore rotates the output by $2\pi/M$.
Writing $z(\phi)=x(\phi)+iy(\phi)$ for the complex output signal:
\begin{equation}
  z\!\left(\phi+\tfrac{1}{M}\right)=e^{\,i2\pi/M}\,z(\phi).
  \label{eq:sym_relation}
\end{equation}
Substituting $z(\phi)=\sum_n\hat{z}_n\,e^{i2\pi n\phi}$
into~\eqref{eq:sym_relation} and equating coefficients by linear
independence of the complex exponentials gives
$\hat{z}_n(e^{i2\pi n/M}-e^{i2\pi/M})=0$,
i.e.\ $\hat{z}_n\bigl(e^{i2\pi(n-1)/M}-1\bigr)=0$,
so $\hat{z}_n\neq0$ only when $n\equiv 1\pmod{M}$.
Since $x(\phi)=\operatorname{Re}(z(\phi))=\tfrac{1}{2}(z+\bar{z})$,
taking the conjugate introduces coefficients at negative indices:
the $n$-th harmonic of $x$ is non-zero also when $n\equiv -1\pmod{M}$.
The active set $n\equiv\pm1\pmod{M}$ defines the \emph{harmonic lattice}:
\begin{equation}
  \mathcal{H}(M)\;=\;\bigl\{(mM\pm 1)\,f_0 \;:\; m=0,1,2,\ldots\bigr\}.
  \label{eq:lattice}
\end{equation}
For $M=N$ this recovers the result of Hohnerlein et al.~\cite{Hohnerlein:2016}
for regular $N$-gons; the derivation here shows it holds for any polygon whose
rotational symmetry order is~$M$, independently of vertex count.
All geometric controls~--- vertex count, edge curvature~$\kappa$, convexity,
and traversal mode~--- leave $\mathcal{H}(M)$ unchanged; arc-length and angular-velocity 
traversal share the same active harmonic set for any $M$-fold symmetric polygon, 
differing only in the amplitudes they assign within the lattice Section~\ref{ssec:traversal_compare}.
$M$ determines \emph{which} harmonics are present;
$\kappa$ and the vertex geometry determine \emph{how much} energy each carries.

\begin{figure}[ht]
  \centering
  \includegraphics[width=1\columnwidth]{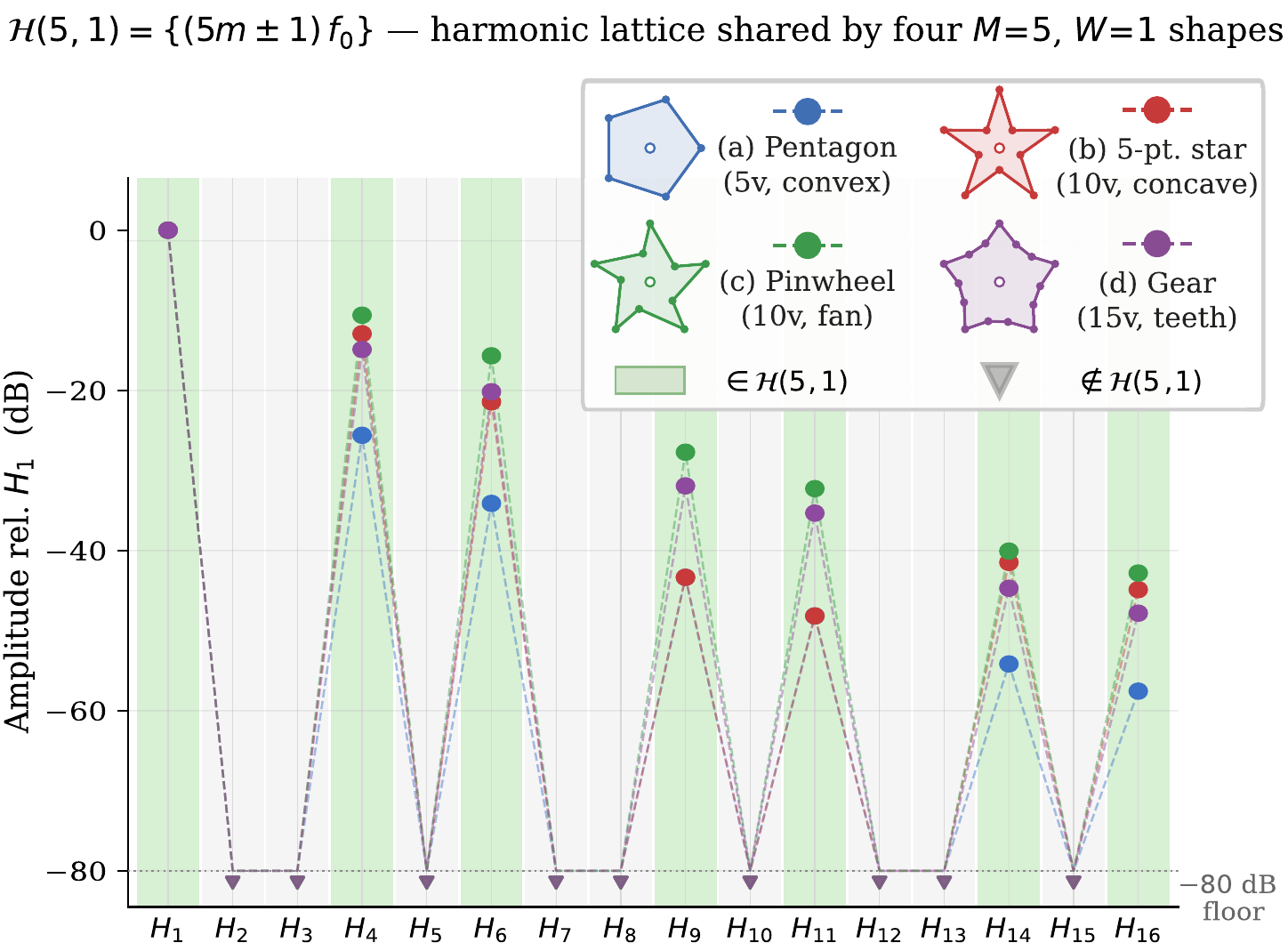}
  \caption{Four $M{=}5$ shapes with different vertex counts share the harmonic
    lattice $\mathcal{H}(5)=\{(5m\pm1)f_0\}$: pentagon (5v, convex),
    five-pointed star (10v, concave), pinwheel (10v, fan), gear (15v, teeth).
    Harmonics outside the lattice fall below the $-80\,\text{dB}$ analysis floor
    regardless of vertex count.
    $f_0=\SI{440}{\hertz}$, $f_s=\SI{48}{\kilo\hertz}$.}
  \label{fig:spectral_characterization_M5}
\end{figure}

\textbf{Extension to self-intersecting polygons.}
Equation~\eqref{eq:lattice} is the $W=1$ case of a more general relation, where
$W$ is the winding number~--- the number of times the traversal encircles the
centroid per period.
The 10-vertex concave star of Section~\ref{ssec:tri_star}
has $W=1$: it is a simple closed curve that winds the centroid
once, so $\mathcal{H}(5,1)$ applies.
When vertices are instead supplied in Schl\"afli order — e.g.\
the pentagram $\{5/2\}$ as 5 vertices winding the centroid
twice per period — advancing the phase by $1/M$ rotates the
output by $2\pi W/M$, giving
$z(\phi+\tfrac{1}{M})=e^{i2\pi W/M}z(\phi)$ and extending the
lattice to
\begin{equation}
  \mathcal{H}(M,W)=\{(mM\pm W)\,f_0\}.
\end{equation}
For $W=2$, $M=5$: $f_0$ is absent and the perceived
fundamental rises to $2f_0$.
\section{Shape Interpolation}
\label{sec:morphing}

Sound morphing is a well-established research area whose common goal is to
obtain gradual timbral transformations between sounds. Established approaches
operate \emph{a posteriori}: a signal is analysed into its constituents and
interpolation is performed in that representation space~--- sinusoidal
partials with unequal feature counts~\cite{Tellman:1995}, the
reassigned bandwidth-enhanced model~\cite{Fitz:2003}, perceptually
motivated descriptors~\cite{Caetano:2013, Kazazis:2016}, or latent
synthesizer parameters regularized by timbre attributes~\cite{LeVaillant:2024}.
Fitz et al.~\cite{Fitz:2003} observe that sound morphing bears a structural
resemblance to geometric morphing in computer graphics, where correspondence
and interpolation are the two complementary problems. The present approach
inverts the direction: rather than deriving geometry from sound, the morphing
trajectory is defined directly in the space of polygon vertices, and the
timbral evolution emerges as a consequence of the geometric interpolation.
As noted by both Caetano and Osaka~\cite{Caetano:2012} and Le~Vaillant and
Dutoit~\cite{LeVaillant:2024}, an interpolation that is linear in a
parameter space is not guaranteed to be perceptually smooth; a perceptual
model of the morphing trajectories is left to future work.
The central technical problem~--- establishing a correspondence between two polygons
with unequal vertex counts~--- is the geometric analogue of the
partial-matching problem; the solution adopted here is described in
Section~\ref{ssec:hybrid}.

\subsection{The Variable Vertex Count Problem}
\label{ssec:morph_problem}

When the polygon is defined by its vertices rather than a formula, the
reachable shapes extend beyond a single parametric family. Where Hohnerlein et
al.~\cite{Hohnerlein:2016} follow one continuous parameter~--- the polygon
order $n$~---tracing a one-dimensional path through shape space (triangle, square, pentagon, and the non-integer forms
between them), a vertex-based oscillator operates in a different region: any closed
polygon (irregular, concave, star-shaped, or freely drawn) can be a source, and
morphing between two such shapes navigates configurations that share no
parametric description.

The challenge is vertex correspondence: interpolating between polygon $A$
($N_A$ vertices) and $B$ ($N_B \neq N_A$) needs a meaningful matching across
unequal sets. Naive linear interpolation is undefined when the counts differ,
and resampling to a common count blurs corner sharpness~--- the very quality
that makes vertex-based synthesis distinctive. The hybrid algorithm below
addresses this while preserving corner sharpness throughout the transition.

\subsection{Hybrid Interpolation Algorithm}
\label{ssec:hybrid}

The expansion strategy is selected per pair of polygons: if both
polygons in a pair are convex, angular correspondence is used for both;
if at least one is concave, perimeter midpoint expansion is applied to
both. This symmetry ensures that the expanded sequences are structurally
compatible and that vertex-to-vertex pairing is meaningful.
Convexity is determined by verifying that the cross products of all
consecutive edge vectors $\overrightarrow{V_iV_{i+1}} \times \overrightarrow{V_{i+1}V_{i+2}}$
share the same sign, which is equivalent to a non-negative signed area
for CCW-ordered vertices.
Self-intersecting polygons are detected by a separate $\mathcal{O}(N^2)$
pairwise edge test: if any two non-adjacent edges cross (verified via
signed-area products), the polygon is additionally classified as
non-convex regardless of cross-product signs.
In both cases the system routes to perimeter midpoint expansion, which
preserves the original vertex order; no intersection points are computed
and the vertex buffer is not modified.
A self-intersecting polygon is therefore traversed faithfully in the
order supplied by the user, yielding crossing output trajectories as
a deliberate timbral option.

\textbf{Both convex~--- angular correspondence.}
Both polygons are expanded to $N_\text{max} = \max(N_A, N_B)$ vertices by
associating each vertex with its angular position relative to the centroid.
For each target angle $\theta_{\text{ref},j}$, the nearest vertex
$\mathbf{V}_{k^*}$ in the source polygon is duplicated:
\begin{equation}
  k^* = \arg\min_k \,\bigl|\theta_k - \theta_{\text{ref},j}\bigr|_{\,2\pi}
  \label{eq:angular_match}
\end{equation}
where $|\cdot|_{2\pi}$ denotes the angular distance modulo $2\pi$.
The duplicated vertices are ``sleeping''~--- they initially coincide with
existing vertices and gradually separate during interpolation, causing new
corners to emerge smoothly.

 For concave polygons, angular correspondence is unreliable: 
concavities cluster multiple vertices into narrow angular sectors as seen from the
centroid, so Eq.~\eqref{eq:angular_match} repeatedly selects the same
source vertex while leaving others unmatched. The resulting sleeping
vertices migrate toward unrelated targets along trajectories that cross
the polygon interior. 

\textbf{At least one concave~--- perimeter midpoint expansion.}
Both polygons are expanded using the same strategy: all original vertices
are preserved in order, and extra vertices are distributed proportionally
to edge length and placed superimposed at edge midpoints:
\begin{align}
  S_i &= \frac{\ell_i}{L}\cdot(N_\text{max} - N),
  \notag\\
  n_{\text{sleep},i} &=
    \begin{cases}
      \lfloor S_i^{\,\Sigma}\rfloor - n_{\text{placed}} & i < N-1,\\
      (N_\text{max}-N) - n_{\text{placed}}              & i = N-1,
    \end{cases}
  \label{eq:midpoint_expansion}\\
  \mathbf{V}_{\text{mid},i} &= \frac{\mathbf{V}_i + \mathbf{V}_{(i+1)\bmod N}}{2}
  \notag
\end{align}
where $S_i^{\,\Sigma} = \sum_{k=0}^{i} S_k$ is the running sum and
$n_{\text{placed}}$ the count of sleeping vertices already assigned.
This Bresenham-style accumulator guarantees that the total equals
$N_\text{max}-N$ exactly, with no separate remainder step.
Superimposed midpoint vertices separate during interpolation,
preserving concavities and characteristic sharpness.
Once both polygons reach equal vertex counts, each matched
pair is interpolated vertex-by-vertex along Cartesian paths.
Before interpolation, the end polygon is cyclically shifted to
the alignment that minimises the total vertex-to-vertex distance:
\begin{equation}
  s^* = \arg\min_{s \in \{0,\ldots,N_\text{max}-1\}}
        \sum_{i=0}^{N_\text{max}-1}
        \bigl\|\mathbf{V}_{A,i} - \mathbf{V}_{B,(i+s)\bmod N_\text{max}}\bigr\|
  \label{eq:cyclic_align}
\end{equation}
This ensures sleeping vertices are paired with geometrically
nearest targets regardless of the original vertex ordering of~$B$.

\subsubsection{Examples}
\label{ssec:tri_star}
\begin{figure}[ht]
  \centering
  \includegraphics[width=1\columnwidth]{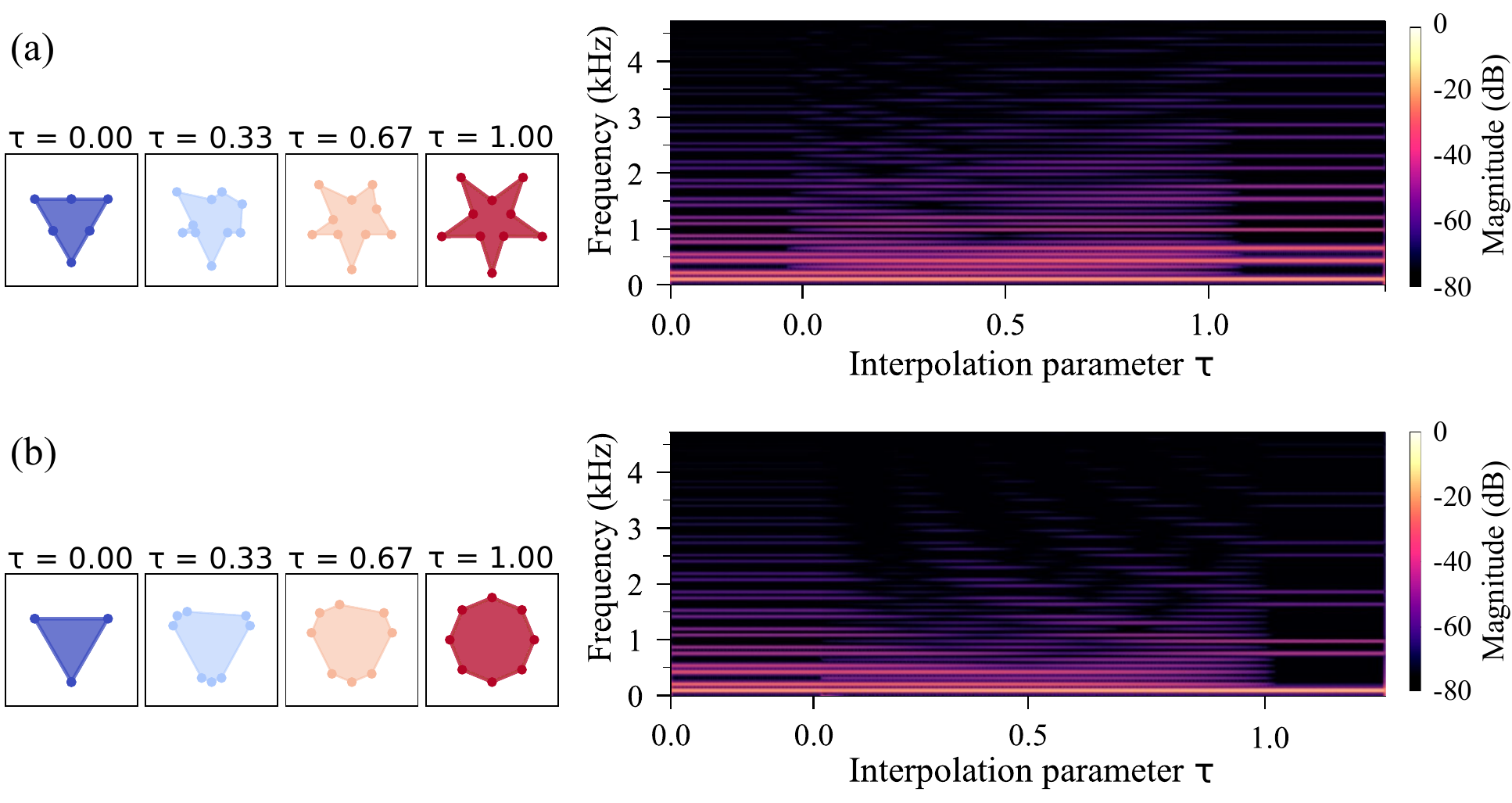}
  \caption{Interpolation strips (left) and magnitude spectrograms over the
    morph (right), 4~frames each.
    (a)~Triangle ($\tau=0$) to five-pointed star ($\tau=1$).
    (b)~Triangle ($\tau=0$) to regular octagon ($\tau=1$).
    $f_0=\SI{110}{\hertz}$, $f_s=\SI{44100}{\hertz}$.}
  \label{fig:tri_morph}
\end{figure}

\textbf{Triangle to Star.} 
The triangle (3~vertices, convex) and the five-pointed star (10~vertices,
concave) trigger perimeter midpoint expansion, since the star is non-convex.
The triangle is expanded to 10~vertices by keeping its three originals and
placing 7~sleeping vertices at edge midpoints, distributed proportionally to
edge length; after cyclic alignment each is paired with one of the star's inner
points. In Figure~\ref{fig:tri_morph} the star tips emerge from the triangle's
vertices while the inner concavities form as the sleeping midpoints migrate
inward.

\textbf{Triangle to Octagon.} The triangle and the regular octagon (8~vertices) are both convex, so angular
correspondence applies to both. The triangle is expanded to 8~vertices by
duplicating its three vertices at the octagon's angular positions (two thrice,
one twice), all sleeping vertices stacked on their source. After cyclic
alignment, each cluster maps to the octagon vertices in the nearest angular
sector: at $\tau=0$ the three clusters sit at the triangle's corners, and as $\tau$
grows they separate toward their octagon targets along straight Cartesian paths
(Fig.~\ref{fig:tri_morph}).

\subsection{Polygon Sequences and Pair Switching}
\label{ssec:pair_switching}

When more than two polygons are defined as a sequence
$P_0,\allowbreak P_1,\allowbreak \ldots,\allowbreak P_n$, each internal polygon $P_k$ acts as a shared
polygon: it serves as the target of pair $(P_{k-1}, P_k)$ and as the
source of the following pair $(P_k, P_{k+1})$. Each pair expands
independently to its own $N_\text{max}$: the left pair uses
$N_\text{max}^{(k-1,k)} = \max(N_{k-1}, N_k)$ and the right pair uses
$N_\text{max}^{(k,k+1)} = \max(N_k, N_{k+1})$, which may differ.
Although the underlying geometry of $P_k$ is identical in both
representations, the sleeping-vertex distribution and vertex ordering
in the buffer differ to satisfy each pair's correspondence requirements,
which would introduce phase jumps and DC offsets at pair boundaries
without corrective measures. Three mechanisms ensure continuity:
boundary phase continuity, vertex $v_0$ consistency via write-back,
and temporal crossfading.

\textbf{Boundary phase continuity.}
Within-pair phase rescaling is suppressed at pair boundaries: since
the two representations of $P_k$ carry slightly different arc-length
perimeters, applying rescaling would shift the traversal position
discontinuously at the switch point.

\textbf{Vertex $v_0$ consistency via write-back.}
We define $v_0$ as the first vertex of the cyclically aligned buffer (position
$s^*$ of Eq.~\eqref{eq:cyclic_align}), the perimeter point where traversal
begins each period at $\phi=0$. For an asymmetric $P_k$ the two expanded representations
may yield different $v_0$ positions, producing a step discontinuity heard as a DC offset. The
write-back step resolves this: once pair $(P_{k-1}, P_k)$ is aligned, the resulting
vertex order of $P_k$ is written back to the \texttt{pbuf} buffer of the
\texttt{PolyManager} overwriting the original user-defined ordering.
This takes place during the geometry build phase, before pair $(P_k, P_{k+1})$
begins~--- so the second expansion inherits the same $v_0$.

\textbf{Temporal crossfading.}
A 20\,ms smoothstep crossfade blends old and new waveform buffers,
holding the morphing parameter $\tau\in[0,1]$ at its boundary value throughout;
polyBLAMP is suspended for its duration (see Section~\ref{sec:aa}).

The selection rules of Section~\ref{ssec:hybrid} propagate via write-back: the
convexity of the written-back polygon is preserved when convex and invalidated
when concave, so angular correspondence resumes at the first pair of two
consecutive convex polygons. Figure~\ref{fig:morph_spectral} shows both strategies
and all three continuity mechanisms; the arrow's asymmetry makes write-back
critical~--- without it a different $v_0$ across pairs~2 and~3 yields a DC
offset.

%
%
%

\subsection{Spectral Structure under Morphing}
\label{ssec:spectral_morph}

The harmonic lattice $\mathcal{H}(M)$ of Section~\ref{ssec:symmetry}
predicts the timbral trajectory of a morph directly from the symmetry
orders of the two endpoint polygons. For a transition from $M_A$ to
$M_B$, three spectral roles are possible for each harmonic~$n$:
harmonics in $\mathcal{H}(M_A)\cap\mathcal{H}(M_B)$ persist throughout;
those exclusive to $\mathcal{H}(M_A)$ fade continuously toward the alias
floor; those exclusive to $\mathcal{H}(M_B)$ emerge from it.
At intermediate~$\tau$ the interpolated shape has no rotational symmetry,
so in principle all integer harmonics are active, but the
out-of-lattice leakage remains small for typical polygon pairs and
the fundamental stays fixed at~$f_0$.

\begin{figure}[ht]
  \centering
  \includegraphics[width=\columnwidth]{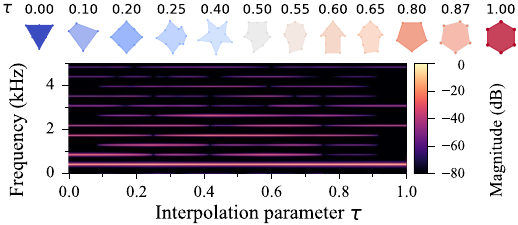}
  \caption{Triangle\,$\to$\,\allowbreak square\,$\to$\,\allowbreak star\,$\to$\,\allowbreak arrow\,$\to$\,\allowbreak pentagon\,$\to$\,\allowbreak hexagon.
    \emph{Top:} shape frames labelled by morph parameter~$\tau$.
    \emph{Bottom:} STFT magnitude over the full sequence. Harmonic lines persist and redistribute while
    the fundamental stays fixed, and no discontinuity appears at the
    pair-switch boundaries.}
  \label{fig:morph_spectral}
\end{figure}

Figure~\ref{fig:morph_spectral} follows a six-shape sequence. Across each
transition the harmonics predicted by the lattice persist, fade, or
emerge as the symmetry order changes, while the fundamental stays
fixed at~$f_0$.

\section{Extension to 3D Polyhedra}
\label{sec:3d}

\subsection{Core Idea: Traversing a Planar Cross-Section}
\label{ssec:3d_core}

The 3D extension changes only the front end. A convex polyhedron is rotated
about its centre by three angles $\theta_x,\theta_y,\theta_z$ and sliced by a
fixed horizontal plane at height $z_\text{plane}$; the resulting convex
cross-section polygon is passed unchanged to the arc-length traversal of
Section~\ref{ssec:traversal} and the polyBLAMP correction of
Section~\ref{sec:aa}. The traversal therefore stays two-dimensional. Only two
elements are specific to 3D: the plane--polyhedron intersection
(Section~\ref{ssec:intersection}) and keeping the output vertex count fixed
across topological transitions, which reuses the sleeping-vertex mechanism of
Section~\ref{ssec:hybrid}. The intersection needs only vertex and face data
and applies to any convex solid; we provide a cube, an icosahedron, and a
square pyramid as case studies.

From a musical standpoint, rotating the solid is a timbral gesture with the
logic of a physical object turning in space: harmonic complexity grows and
recedes at the topological transitions where the plane crosses a vertex or
edge, giving natural points of articulation within a continuous sweep.
Because the oscillator output is natively two-dimensional, spatial
orientation maps directly onto stereo movement, the $x$/$y$ phase
relationship evolving with the rotation.

\subsection{Polyhedron Intersection Algorithm}
\label{ssec:intersection}

After the rotation is applied to all vertices, each edge
$(\mathbf{V}_i,\mathbf{V}_{i+1})$ is tested against the plane through the
signed distance $d_i = z_i - z_\text{plane}$. An edge crosses when $d_i$ and
$d_{i+1}$ differ in sign, at the point
\begin{equation}
  \mathbf{P}_\text{int} = \mathbf{V}_i + \alpha\,(\mathbf{V}_{i+1}-\mathbf{V}_i),
  \qquad \alpha = \frac{-d_i}{\,d_{i+1}-d_i\,}.
  \label{eq:edge_intersect}
\end{equation}
Edges coplanar with the plane ($|d_{i+1}-d_i|<\varepsilon$) are skipped to
avoid a $0/0$ indeterminacy. The crossing points are sorted by angle about
their arithmetic-mean centroid to form the cross-section polygon, with
duplicates merged when the plane passes through a polyhedron vertex.

\subsection{Topological Stability via Sleeping Vertices}
\label{ssec:sleeping_vertex}

When the plane crosses a polyhedron vertex during rotation, the number of
real intersection points changes by $\pm1$ or $\pm2$; left unhandled, the
output vertex count would change instantaneously and disrupt both the
traversal and the polyBLAMP correction. We avoid this with the
sleeping-vertex scheme of Section~\ref{ssec:hybrid}, here applied to
cross-section transitions: the output count is fixed at the edge count $n_e$
of the solid, and every non-crossing edge contributes a \emph{sleeping
vertex} co-located with a real one, producing a zero-length edge that the
arc-length traversal skips.

The intersection runs in two passes. Pass~1 collects the
$n_\text{real}\ge3$ real crossing points via
Eq.~\eqref{eq:edge_intersect} and sorts them angularly. Pass~2 parks each
sleeping vertex: with reference position $\mathbf{s}_j$ the 2-D projection of
the edge endpoint nearer the plane,
\begin{equation}
  \mathbf{s}_j = \begin{cases}
    (x_{V_i},\, y_{V_i})         & |d_i| \le |d_{i+1}| \\
    (x_{V_{i+1}},\, y_{V_{i+1}}) & \text{otherwise,}
  \end{cases}
  \label{eq:sleep_ref}
\end{equation}
the vertex is parked at the nearest real vertex,
\begin{equation}
  \mathbf{P}_\text{sleep} = \underset{\mathbf{P}_k}{\arg\min}\;
    \bigl\|\mathbf{s}_j - \mathbf{P}_k\bigr\|^2,
  \label{eq:sleep_park}
\end{equation}
and inserted immediately after its host, guaranteeing a zero-length edge.
The total count $n_e = n_\text{real} + n_\text{sleep}$ is constant, so the
engine never sees a vertex-count change and the polyBLAMP correction stays
active throughout. Continuity is $C^0$: as the plane approaches a
non-crossing edge, $\alpha\to0$ and $\mathbf{P}_\text{int}\to\mathbf{V}_i$,
already coinciding with the sleeping position.

\subsection{Case Studies}
\label{ssec:case_studies}

\begin{figure}[ht]
  \centerline{\includegraphics[width=.68\columnwidth]{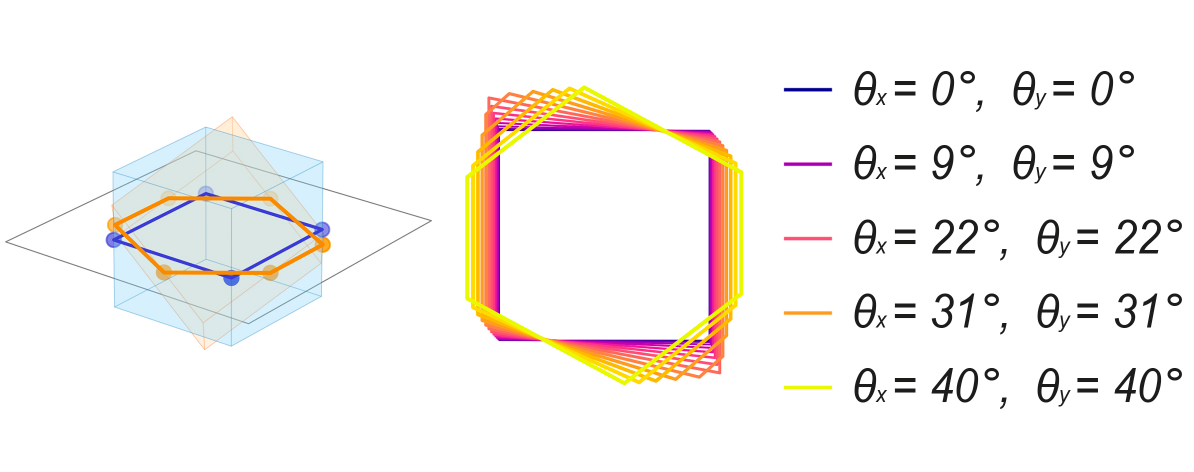}}
  \caption{Cube cross-section from square ($\theta_x=\theta_y=0^\circ$, blue)
    to near-regular hexagon ($\theta_x=\theta_y=40^\circ$, amber).}
  \label{fig:cube_full}
\end{figure}

Rotating a \textbf{cube} sliced at $z_\text{plane}=0.5$ sweeps the
cross-section from a square to a regular hexagon
(Fig.~\ref{fig:cube_full}), the vertex count moving between 4 and 6.
A \textbf{square pyramid} sliced at the same height sweeps from a square
($\theta_x=0^\circ$) to a triangle ($\theta_x=90^\circ$) through
intermediate irregular quadrilaterals as the apex enters the section,
reducing the count from 4 to 3. Both yield a continuous shift in harmonic
content; companion page.\footnote{\url{https://www.antonioargentieri.com/polygon_demo/}}

\section{Signal Quality and Practical Considerations}
\label{sec:aa}

At each vertex the traversal direction changes abruptly, giving a
first-derivative discontinuity and an unbounded spectrum that aliases when
sampled. Hohnerlein, Rest, and Parker~\cite{parker2017} addressed this for the
continuous-order parametric oscillator by deriving at each vertex a
closed-form derivative jump, enabling four-point polyBLAMP correction.
For arbitrary vertex buffers that per-shape derivation no
longer applies, as there is no parametric family; the jump, however, still
admits a closed form, obtained directly from the tangent directions of the two
adjacent Bézier edges at their common vertex~--- valid for any vertex
configuration.
This motivates the strategy adopted here: a four-point polyBLAMP correction
from the local geometry, combined with adaptive oversampling.

\subsection{Antialiasing Strategy}

\subsubsection{Four-Point polyBLAMP Correction}
\label{ssec:polyblamp}

At each vertex $\mathbf{V}_i$ the arc-length traversal transitions from the incoming
quadratic Bézier edge to the outgoing one.
The traversal signal is the two-dimensional coordinate $\mathbf{P}(\phi) = (p_x, p_y)$
of the point on the polygon at phase $\phi \in [0,1)$. Each component undergoes
a derivative discontinuity at $\phi = \phi_i = s_i / L$, where $s_i$ is
the cumulated arc length to $\mathbf{V}_i$ and $L$ is the total perimeter.

The outgoing and incoming tangent directions at $\mathbf{V}_i$ are the
exact endpoint tangents of the quadratic Bézier, normalised to
arc-length units (for $\kappa=0$ they reduce to the chord directions
$(\mathbf{V}_{i+1}-\mathbf{V}_i)/\ell_i$):
\begin{align}
  \mathbf{D}^+_i &= \frac{2\,(\mathbf{P}_{c,i} - \mathbf{V}_i)}{\ell_i}
  \label{eq:dout}\\
  \mathbf{D}^-_i &= \frac{2\,(\mathbf{V}_i - \mathbf{P}_{c,i-1})}{\ell_{i-1}}
  \label{eq:din}
\end{align}
These follow from the derivative of the quadratic Bézier
$\mathbf{B}(u) = (1-u)^2 \mathbf{V}_i + 2(1-u)\,u\,\mathbf{P}_c + u^2 \mathbf{V}_{i+1}$,
whose derivative is
$\mathbf{B}'(u) = 2(1-u)(\mathbf{P}_c - \mathbf{V}_i) + 2u(\mathbf{V}_{i+1} - \mathbf{P}_c)$,
giving $\mathbf{B}'(0) = 2(\mathbf{P}_c - \mathbf{V}_i)$ and $\mathbf{B}'(1) = 2(\mathbf{V}_{i+1} - \mathbf{P}_c)$.
Dividing by $\ell_i$ (resp.\ $\ell_{i-1}$) normalises to arc-length units.
When $\kappa = 0$, $\mathbf{P}_{c,i} = (\mathbf{V}_i + \mathbf{V}_{i+1})/2$, so
$\mathbf{D}^+_i = (\mathbf{V}_{i+1} - \mathbf{V}_i)/\ell_i$ and
$\mathbf{D}^-_i = (\mathbf{V}_i - \mathbf{V}_{i-1})/\ell_{i-1}$~--- the chord directions~---
recovering the piecewise-linear case as a special instance of the general formula.

The derivative jump vector
$\Delta\mathbf{D} = \mathbf{D}^+_i - \mathbf{D}^-_i$
is in units of $[\text{coordinate} / \text{arc length}]$.
Scaling by the arc distance covered per oversampled sample
$L \cdot \Delta\phi$, where $\Delta\phi = f_0 / f_s^{\mathrm{OS}}$, converts
it to signal change per sample:
\begin{equation}
  \mathbf{J} = \Delta\mathbf{D} \cdot L \cdot \Delta\phi
  \label{eq:jump}
\end{equation}
$\mathbf{J} = (J_x, J_y)$ captures the magnitude and sign of the
derivative jump for each cartesian component independently.

The fractional delay $d \in [0,1)$ locates the vertex within the
current oversampled sample:
\begin{equation}
  d = \frac{\phi - \phi_i}{\Delta\phi}
  \label{eq:d}
\end{equation}
where the condition $d \in [0,1)$ is enforced by detecting the single
oversampled sample immediately following the vertex crossing
($\phi - \phi_i \in [0,\,\Delta\phi)$, with $\phi$ and $\phi_i$ both
in $[0,1)$).

Applying the four-point polyBLAMP residuals
of~\cite{parker2017,esqueda2016} (Table~1 therein),
$r_{-2}$, $r_{-1}$, $r_0$, $r_{+1}$, scaled by $\mathbf{J}$, to the four samples surrounding the
discontinuity ($r_{-2}$: two steps before, buffer \texttt{prevprev};
$r_{-1}$: one step before, \texttt{prev}; $r_0$: one step after,
\texttt{acc}$_1$; $r_{+1}$: two steps after, \texttt{acc}$_2$):
\begin{align}
  \texttt{prevprev} &\mathrel{+}= \mathbf{J}\, r_{-2}(d)
  \label{eq:acc_prevprev} \\
  \texttt{prev}   &\mathrel{+}= \mathbf{J}\, r_{-1}(d)
  \label{eq:acc_prev} \\
  \texttt{acc}_1  &\mathrel{+}= \mathbf{J}\, r_{\phantom{+}0}(d)
  \label{eq:acc1} \\
  \texttt{acc}_2  &\mathrel{+}= \mathbf{J}\, r_{+1}(d)
  \label{eq:acc2}
\end{align}
The sign of $\mathbf{J}$ encodes the direction of the derivative jump, so
addition here is equivalent to the subtraction convention
of~\cite{parker2017} where a fixed-sign jump is used.

At each oversampled step the output is \texttt{prevprev}; the buffer advances as
$\texttt{prevprev} \leftarrow \texttt{prev}$,
$\texttt{prev} \leftarrow \mathbf{P} + \texttt{acc}_1$,
$\texttt{acc}_1 \leftarrow \texttt{acc}_2$, $\texttt{acc}_2 \leftarrow 0$,
where $\mathbf{P} = \mathbf{P}(\phi) = (p_x, p_y)$ is the uncorrected oscillator sample
at the current oversampled phase, obtained directly from the
arc-length traversal of Eq.~\eqref{eq:bezier_traversal} before any
polyBLAMP residual is applied.
This introduces a latency of two oversampled samples ($2/f_s^{\mathrm{OS}}$),
below the audible threshold at all oversampling factors used here.

The correction is skipped when either the current or the previous edge length
falls below a threshold ($\ell < 10^{-3}$), guarding against division by zero
for degenerate zero-length edges (sleeping vertices in star-polygon configurations).

The correction is suspended during pair-switch crossfades: although the
shared polygon is geometrically identical at the boundary, the change in
$n_\text{vMax}$ between adjacent pairs redistributes the arc-length phase
across a different number of slots, rendering the accumulated polyBLAMP
residuals incoherent with the new buffer layout.

\subsubsection{Adaptive Oversampling}

The polyBLAMP correction alone reduces the alias floor by approximately
22\,dB relative to the uncorrected signal,
as confirmed by the measurements below. Residual aliasing is further
attenuated by adaptive oversampling:
\begin{equation}
\text{OS} =\mathrm{clamp}\!\left(\left\lfloor\frac{44100}{f_s} \cdot\bigl(2 + 4\,f_\text{norm}\bigr)\right\rfloor,\;2,\;6\right)
\end{equation}
where $f_\text{norm} = \sqrt{\mathrm{clamp}(f_0 - 200, 0, 7800)/7800}$. The
empirically chosen bounds $[\SI{200}{\hertz},\,\SI{8000}{\hertz}]$
bracket the range where aliasing is perceptually relevant:
below \SI{200}{\hertz} OS\,=\,2 is already sufficient, and above
\SI{8000}{\hertz} additional oversampling yields negligible benefit. The factor $44100/f_s$ scales
the rate inversely with sample rate (at $f_s = 88200$\,Hz it halves the OS,
the higher native rate already pushing alias products out of range), and the
sub-linear $f_\text{norm}\propto f_0^{1/2}$ reflects the diminishing benefit at
high $f_0$. A critically-damped ($Q=0.5$) low-pass biquad at
\SI{20}{\kilo\hertz}~\cite{RBJ:cookbook} removes residual high-frequency
content before decimation.

\subsubsection{Measured Performance}
\label{ssec:snr}

Table~\ref{tab:snr} reports the measured SNR
for an equilateral triangle (3-vertex buffer, $f_s = 44100$\,Hz) under four
configurations (no antialiasing; OS\,=\,2 only; polyBLAMP only, OS\,=\,1;
OS\,=\,2 with polyBLAMP), at edge curvature $\kappa = 0$ and
$\kappa = -0.234$. SNR is defined as
\begin{equation}
  \mathrm{SNR} = 10\log_{10}\!\frac{P_\text{harm}}{P_\text{alias}}
  \label{eq:snr}
\end{equation}
where $P_\text{harm}$ is the energy within $\pm 4$\,FFT bins of each
harmonic $kf_0$, $k = 1,\ldots,\lfloor f_s/(2f_0)\rfloor$, computed from
a 4\,s Blackman-windowed segment of the output signal normalised to 0\,dBFS
(FFT of $N = 176\,400$ points, bin spacing $0.25$\,Hz);
$P_\text{alias}$ is the remaining spectral energy.

\begin{table}[ht]
\caption{SNR (dB) for an equilateral triangle (3-vertex buffer,
         $f_s = 44100$\,Hz). Best value per row in bold.}
\centering
\label{tab:snr}
\begin{tabular}{l r r r r}
\toprule
$f_0$ & No AA & OS\,=\,2 & BL4 only & OS\,=\,2+BL4 \\
\midrule
\multicolumn{5}{l}{\textit{Straight edges ($\kappa=0$)}} \\
400\,Hz  & 58.8 & 77.6 & 80.9 & \textbf{85.1} \\
751\,Hz  & 51.1 & 75.6 & 75.3 & \textbf{79.1} \\
1350\,Hz & 43.9 & 67.7 & 66.2 & \textbf{70.1} \\
\midrule
\multicolumn{5}{l}{\textit{Edge curvature ($\kappa=-0.234$)}} \\
400\,Hz  & 57.8 & 78.5 & 79.3 & \textbf{83.4} \\
751\,Hz  & 49.5 & 73.9 & 73.7 & \textbf{77.4} \\
1350\,Hz & 42.1 & 66.3 & 64.0 & \textbf{68.3} \\
\bottomrule
\end{tabular}
\end{table}

The BL4 gain over No\,AA ($\approx$\,22\,dB) is consistent across both
curvature conditions, confirming that the Bézier tangent computation of
Eqs.~\eqref{eq:dout}--\eqref{eq:din} preserves polyBLAMP effectiveness for
curved edges. For $\kappa = 0$, BL4 alone exceeds OS\,=\,2 at 400\,Hz, while at
751 and 1350\,Hz the two are comparable.
The lower SNR at higher frequencies reflects the greater
aliasing in the original signal, not a degraded correction: the fractional
delay is known exactly regardless of $f_0$, so polyBLAMP stays fully effective,
but the alias products are then more numerous and closer to the harmonics,
making broadband oversampling relatively more effective.
OS\,=\,2+BL4 achieves the highest SNR in all conditions, confirming the two are
complementary: polyBLAMP targets leakage at the discontinuities, oversampling
provides broadband suppression independent of geometry. Only the final filtered sample per audio-rate block is written to the output.


\section{Implementation}
\label{sec:implementation}

The system is implemented as two RNBO (Cycling~'74)
\texttt{codebox\char`\~} modules: a \texttt{PolyManager} that stores the polygons and builds the
interpolation pairs, and a DSP engine running the per-sample traversal,
polyBLAMP and crossfade. The full implementation is
available.\footnote{\url{https://github.com/antonioargentieri1/Arbitrary_Polygon_Oscillator}}

\textbf{Geometry caching.} The geometry pipeline~--- centroid, vertex
transformation, control-point placement, edge-length and perimeter
accumulation~--- runs only on a parameter change beyond a small tolerance, not
per sample; the phase is then rescaled to the new perimeter to hold pitch,
suppressed at pair-switch boundaries where the crossfade of
Section~\ref{ssec:pair_switching} maintains continuity.

\textbf{Computational cost.} Pair building is $\mathcal{O}(P N^2)$ for $P$ pairs
of up to $N$ vertices~--- the self-intersection test, the angular sort and
nearest-angle expansion, and the cyclic $v_0$ alignment each cost
$\mathcal{O}(N^2)$, the remaining stages $\mathcal{O}(N)$~--- but with
$N\!\le\!24$ and $P\!\le\!7$ it amounts to a few thousand operations, run once
when the pair set is rebuilt rather than per sample. The audio path is then only
$\mathcal{O}(\text{OS}\cdot N)$, $\text{OS}\!\in\![2,6]$: one $\mathcal{O}(N)$
arc-length edge search with $\mathcal{O}(1)$ polyBLAMP per oversampled step, over
fixed, pre-allocated buffers with no audio-thread allocation.


\section{Conclusions}
\label{sec:conclusions}

The arc-length engine generalises polygonal synthesis to arbitrary
shapes, morphable sequences, and 3D polyhedral cross-sections, with the
sleeping-vertex algorithm ensuring topological stability across all transitions.
The central shift is from shape as a parameter to shape as a drawing.
Future work centres on perceptually guided morphing and higher-order polyhedra.
\bibliographystyle{IEEEtranDAFx}
\bibliography{Polygon_DAFx26}

\end{document}